# The Ecology of Fringe Science and its Bearing on Policy

**Harry Collins, Andrew Bartlett and Luis Reyes-Galindo**

School of Social Sciences, Cardiff University[1]

emails: CollinsHM@cardiff.ac.uk, luisreyes@ciencias.unam.mx, BartlettA@cardiff.ac.uk

## Introduction

Fringe science has been an important topic since the start of the revolution in the social studies of science that occurred in the early 1970s.[2] As a softer-edged model of the sciences developed, fringe science was a 'hard case' on which to hammer out the idea that scientific truth was whatever came to count as scientific truth: scientific truth emerged from social closure. The job of those studying fringe science was to recapture the rationality of its proponents, showing how, in terms of the procedures of science, they *could* be right and the mainstream *could* be wrong and therefore the consensus position is formed by social agreement.

One outcome of this way of thinking is that sociologists of science informed by the perspective outlined above find themselves short of argumentative resources for demarcating science from non-science. The distinction with traditional philosophy of science, which readily demarcates fringe subjects such as parapsychology by referring to their 'irrationality' or some such, is marked.[3] For the sociologist of scientific knowledge, that kind of demarcation comprises

---

[1] This paper is joint work by researchers supported by two grants: ESRC to Harry Collins, (RES/K006401/1) £277,184, *What is scientific consensus for policy? Heartlands and hinterlands of physics* (2014-2016); British Academy Post-Doctoral Fellowship to Luis Reyes-Galindo, (PF130024) £223,732, *The social boundaries of scientific knowledge: a case study of 'green' Open Access* (2013-2016). The second of these projects was initially based on the thinking that inspired the first. Andrew Bartlett is the full-time researcher on the first project. Interviews with Paul Ginsparg and most of the research on arXiv and viXra were conducted by Reyes-Galindo; nearly all the ongoing fieldwork on the 'beyond-arXiv' fringe was conducted by Bartlett, including 17 interviews with active participants in fringe physics whose work is based in the UK, USA, and Australia and ethnographic observation at the 2014 Natural Philosophy Alliance conference in Baltimore, the 2015 Electric Universe conference in Phoenix, and the 2015 Chappell Natural Philosophy Society conference in Boca Raton. The paper has been greatly improved following discussion at the weekly seminar of the Centre for the Study of Knowledge, Expertise and Science (KES).

[2] This is what Collins and Evans (2002) refer to as the 'Second Wave of Science Studies' while this paper could be said to be an exercise in 'Third Wave Science Studies.'

[3] A recent collection that attempted to revisit the problem of demarcation is Pigliucci and Boudry's *Philosophy of Pseudoscience* (2013), contains, in its introduction the line "we purposefully steered clear from the kind of sociology inspired by social constructivism and postmodernism – which we regard as a type of pseudodiscipline in its own right" (p4). Not

a retrospective drawing on what is found within the scientific community. In contrast, the sociological perspective explains why a multiplicity of conflicting views on the same topic, each with its own scientific justification, can coexist. A position that can emerge from this perspective is to argue for less authoritarian control of new scientific initiatives – for a loosening of the controls on the restrictive side of what Kuhn (1959, 1977) called 'the essential tension'. The essential tension is between those who believe that science can only progress within consensual 'ways of going on' which restrict the range of questions that can be asked, the ways of asking and answering them and the kinds of criticism that it is legitimate to offer – this is sometime known as working within 'paradigms' – and those who believe that this kind of control is unacceptably authoritarian and that good science is always maximally creative and has no bounds in these respects. This tension is central to what we argue here. We note only that a complete loosening of control would lead to the dissolution of science. What drives this paper is a different question that arises out of the sociological perspective: What is the outside world to do with the new view?

This question is *essentially* a policy question even though the cases we will deal with are not policy questions in the conventional sense. This question is not about how science should be conducted, it is about what decision-makers should do with the multiplicity of conflicting views that are found within science itself. Consider the recent detection of gravitational waves (Abbott, B. P. *et al.* 2016, 'Observation of Gravitational Waves from a Binary Black Hole Merger' *Physical Review Letters*, 116, 061102)[4]: while the mainstream scientific community exhibited a quite remarkable consensus over the soundness of the detection, within days there were criticisms and alternative interpretations emerging from the fringe including a lively blog promulgated by Nature and at least two full papers promulgated on the 'alternative' physics pre-print server 'viXra' (viXra:1603.0127; viXra:1603.0232).[5] Mainstream scientists and scientific policymakers ignore such things, but to know what to take seriously and what to ignore requires that one understand the way the scientific community works – one must possess what we call 'Domain Specific Discrimination' ('Domain Discrimination' for short) which is a component of specialist tacit knowledge (Collins, Ginsparg and Reyes-Galindo, forthcoming).[6]

But what are sociologists working under the prescriptions of Wave 2 of science studies to make of these alternative claims and, more to the point as far as this paper is concerned, what is to be made of them by those who run funding agencies? After all, if the alternative accounts are correct, then the roughly *billion dollars* spent on the detection of gravitational waves has been wasted and the next billion being demanded to exploit the discovery and develop a new

---

surprisingly, the section of the book titled 'The History and Sociology of Pseudoscience' was light on any serious, contemporary sociology. This paper takes the insights of Science Studies seriously. See Van Rillaer (1991) and Park (2000) for typical scientist/skeptic rationalist accounts of 'mistaken' and 'irrational' science and of 'pseudoscience'.

[4] A book length account of this discovery will be published as Collins (2017).

[5] See below for more on viXra

[6] For specialist tacit knowledge see Collins and Evans, (2007); for Domain Specific Discrimination see Collins and Weinel (2011).

gravitational wave astronomy would be wasted too. What will happen, of course, is that the funders and other policy-makers will follow the lead of the mainstream but surely it behoves social analysts of science to say something more about the relationship between the mainstream and the fringe, not just leave things to work out as they will. We do believe that it is proper in democratic societies for policy-makers and other decision-makers to begin their technological decision-making work from the consensus of the mainstream, but we want to do a better job of saying what this means.

Here we try to begin a program of research that will lead to a better understanding of the relationship between mainstreams and fringe and, since we want our theory to be general, not tied to specific policy questions such as the safety of tobacco or global warming where the right way to jump is 'over-determined': we will concentrate on problems without an immediate policy relevance beyond questions of funding – a kind of hard case. We will look at physics.

We take it that all the physicists we discuss here, including the fringe and independent physicists, just want to do physics better, as they see it. Many of these physicists are often well-qualified in their own disciplines, hold university posts and publishing now or published at an earlier stage in their careers in mainstream physics journals. The problem we deal with here is closely related to what Collins and Evans (2002) called 'the problem of extension', which is to do with demarcation of the use of the term 'expert'.

**The ecology of the fringe**

The paper begins by trying to provide an 'ecology' of the hinterlands of physics, a description of those who live there and how their activities compare with the mainstream as well as the ebbs and flows between the permeable boundaries of consensus and dissent.[7] We concentrate on physics but will mention other sciences in passing. We provide a descriptive overview of the problem of demarcation as encountered in physics all the way from the heartlands to the hinterlands and on to the outer reaches but in this paper our analytic focus will be on the hinterlands and outer reaches only; for an analysis of the heartlands and the 'marshy ground' at the borders of physics see Reyes-Galindo (2016).

The paper can be thought of as divided into two main parts. The first part *describes* the heartlands and marshy ground and provides the background for the second part which completes the description and also *analyses* the outer reaches. This analysis indicates the first-principles of a style of thinking which we hope, will be applied to the, much more difficult to deal with, heartlands and marshy ground in future research.

*Heartlands*

We begin the description in the heartlands of physics, concentrating on the open access electronic preprint server, 'arXiv', where demarcation has presented itself as a day-to-day problem from its foundation. The importance of the arXiv is due to its being one of the main sites for the dissemination and acquisition of up-to-date physics knowledge, in some disciplines being far more important than traditional, peer-reviewed journals. As we will explain in greater detail, arXiv exiles some submissions to a low status 'General Physics' category. What it means to be posted in General Physics is understood within the physics community but the nature of the demarcation is not described on the server's own site or in its literature (Reyes-Galindo 2016). A member of the public, or a politician, trying to educate themselves by downloading physics preprints from arXiv would be quite unaware that a paper in General Physics has to be treated differently to papers in other categories. The point is that But the meaning of a paper cannot be acquired from its words and symbols in the absence of an understanding of its social setting because nearly all the papers we discuss here are written by active physicists and look like regular physics papers.[8]

---

[7] Delborne (2008) has proposed five 'key boundaries' in science to understand dissent science, two of which are relevant to our analysis (see below).

**Physics and arXiv**

Physics has a long history of dealing with the problem of the fringe and a rich hinterland of fringe groups; though it must still draw boundaries, physics is also relatively tolerant of maverick ideas (Kaiser 2011). Of the inner end of the physics fringe, Michael Berry, former editor of the *Proceedings of the Royal Society* remarked:

> With a journal of such prestige we get a lot of junk, people who aren't scientists with a new theory. Often retired engineers seem to be prone to this grandiosity. You instantly know if a paper is junk, but on the other hand you have to take into account that the author is serious, and has thought a great deal about what they've done.[9]

Physicists have tried to resolve the problem themselves by characterising the special nature of fringe science: Baez (1998), Siegel (2011) and 't Hooft (2003) are attempts to define outsiders which adopt a waggish, 'jokey' style, perhaps to relieve the stress of the surgical exercise when they cannot be sure to the standards of logic that the 'organs' being discarded are not healthy;[10] Langmuir (1989) is more serious.[11]

---

[8] As with 'Primary Source Knowledge' (Collins and Evans 2007)

[9] Interview by Luis Reyes-Galindo, 6 April, 2010. Martin Gardner (1957, p. 8), well-known *Scientific American* columnist and arch-sceptic of all things unorthodox, wrote about the 'illegitimate' world of physics as full of 'stupid, ignorant, almost illiterate men who confine their activities to sending 'crank letters' to prominent scientists', but acknowledged that others are 'brilliant and well-educated, often with an excellent understanding of the branch of science in which they are speculating.'

[10] See Becker et al (1961). A less benevolent reading is suggested by Thérèse and Martin (2010) who point out that this style of satirical 'public shaming' exercises are examples of 'degradation rituals' familiar to sociologists.

[11] The normal process of socialisation into a profession, of course, brings with it a tacit sense of what is to be taken seriously. When forced to reflect, gravitational wave scientists provided Collins (2014) with the following justifications for ignoring a published paper which questioned the basis of their work: tacit aspects of style; never heard of the journal; never heard of the

For arXiv, boundary work cannot be avoided.[12] arXiv was set up in 1991 as an electronic preprint server for physics based on a long standing practice of preprint dissemination in high-energy theory.[13] It is now the preferred means of promulgating findings in many areas of physics, the peer-reviewed journals being too slow to keep up with the moving frontiers of research. According to Larivière et al (2014), approximately 70%-80% of all papers published in physics journals are first posted to arXiv (specific percentages vary across different specialties) with peer-reviewed journals being assigned the roles of *certification* and long-term *preservation*.[14] arXiv was, from its beginning, intended as a dissemination channel for 'professional' physicists only, as founder Paul Ginsparg explained:

> Right from the outset 20 years ago, the guiding principle was that we were trying to provide a resource for research professionals to communicate to one another and that it was not supposed to be a megaphone for people outside the research community to broadcast into the community. I phrase it in those terms because the most frequent issue that we had in the period from the mid-90s to the early 2000s was that people regarded it as the analogue of a Usenet newsgroup… which was this open, free-for-all, Jeffersonian democracy. People were confused, 'oh, it's on the Internet, it's supposed to be open! It's completely open.'… It was designed so that anybody in the amateur community can read it, but this is not a mechanism for you to expose your exciting theories, especially if you haven't had an ordinary research training.[15]

arXiv currently supports 13 specialist physics domains with subject-specific sub-domains. There are five other domains supporting individual areas of other kinds of science.[16] Across all categories arXiv currently stores over one million accessible articles with around 8,000 new articles being posted every month. According to arXiv's founder, Paul Ginsparg (2003), 'from the outset, a variety of heuristic screening mechanisms have been in place [in arXiv] to ensure insofar as possible that submissions are at least of refereeable quality. That means they satisfy the

---

author; never come across this article or similar by this author; author has little record of scientific accomplishment; journal and paper are incestuous in terms of author list and citation pattern; typical cranky anti-relativity paper and anti-relativity is past its sell-by date.

[12] For 'boundary work' see Gieryn (1983, 1999). For details of boundary work within arXiv physics categories, see Reyes-Galindo (2016).

[13] See Ginsparg, (1994, 2011) and Kaiser, (2005).

[14] arXiv currently registers over 10,000,000 downloads per month, with 573,119,257 total downloads as of February 2015. arXiv, however, does *not* publish statistics on individual paper downloads. "Frequently Asked Questions on Public Statistics", URL: http://arxiv.org/help/faq/statfaq

[15] Interview with Reyes-Galindo 16 June, 2014.

[16] Mathematics, computer science, quantitative biology, quantitative finance, and statistics.

minimal criterion that they would not be peremptorily rejected by any competent journal editor as nutty, offensive, or otherwise manifestly inappropriate, and would instead at least in principle be suitable for review.'  Nevertheless, the number of what were considered to be unsuitable postings has caused arXiv to erect tougher hurdles which have (a) made it harder to post and (b) classified the postings in ways to provide an indication of which, in the moderators' view, should be taken less seriously. The hurdles include:

*The 'reclassification' criterion*:  In October 1996 the general physics (*physics.gen-ph*) subcategory was established into which could be diverted postings from professional physicists which were, nevertheless, considered to be less worthy of serious consideration than arXiv's staple diet. Ginsparg explained this category as follows:

> The American Physical Society had long had an issue at their Annual Meetings … They would get abstracts from all over and anybody could become a member –roughly speaking– and they had a commitment to let the members present and so they invented this notion of the 'general physics session' at their meetings where the stuff, the abstracts and the presentations that none of the 'real' sections wanted would be shunted to the general physics section … We usually employed it for people who at least came from a conventional background. It wouldn't be for [non-papers] [o]r things without references; or things that manifestly violate known-principles; high school students refuting special relativity; perpetual motion machines; cold-fusion and all the rest ... So people who had conventional associations or some kind of past publication record, who had either momentarily or permanently gone off in some direction that was no longer of interest to physicists, we would put in the general physics category.[17]

Considering physics alone, arXiv currently posts around 3,500 papers per month into its specialist sections with around 30 papers per month being classified into the 'general physics' subdomain.[18]  Authors have sometimes found themselves excluded from arXiv altogether or permanently diverted to the *gen-ph* subcategory after an early career in which they were able to post in the specialist sections.  Such transitions are perceived to have, and probably do have, significant consequences for careers and for credibility and there are many discussions on the web which describe the perceived injustices of the treatment (Reyes-Galindo 2016).[19]

*The 'nominal refereeabilty' criterion*:  In the mid to late 90s, arXiv moderators were asked to carry out their task by 'embodying' the prototype of a researcher in their community and only allow postings that would be of minimal interest to their research community, and, peremptorily, to reject the rest.  Ginsparg explained:

---

[17]  Interview with Reyes-Galindo 16 June, 2014.

[18]  "arXiv monthly submission rates", URL: http://arxiv.org/stats/monthly_submissions

[19]  See for example, "The Consequences of Being Blacklisted from Posting to ArXiv.org", URL: http://www.archivefreedom.org/freedom/consequences.html

It was also in the mid- to late-90s [that] I was in discussions with the American Physical Society's Editor in Chief and got a better feeling for the way their operations worked. I employed this notion of 'refereeable' in the sense that a competent editor-in-chief of a journal, when receiving an article, will make a spot decision, 'is this worth sending out to referees, or should it be' - and these are the words that Marty used – 'peremptorily rejected by the editor as not of interest to the community?' Because they have this decision criterion, which is their referees are a scarce resource and you don't want to waste their time by sending them stuff that they just look at and just drop. 'This isn't worth my time. Why are you bothering me with sending me this?' […] That was the 'nominal refereeability criterion', that if a moderator would look at it and say 'if this went to a journal it wouldn't even be sent to a referee' we would not want it for the main section.[20]

*The 'endorsement' criterion*: Introduced in January 17, 2004, the endorsement criterion is that all first-time posters be sponsored by two scientists already endorsed through their posting record. Additionally, automatic affiliation can be given 'based on topic, previous submissions, and academic affiliation.'[21] Ginsparg remarked:

My criterion, based on training and background and publication record, is that we regard as the community the people we know and respect and the people they refer to us. You know, our friends and the friends of our friends in this generalised sense and that defines the community.[22]

Endorsement is also tied to (re)classification: authors can only endorse to categories that they themselves are endorsed for. Endorsers must additionally be active in the category, or lose endorsement privileges.

A final obstacle to ready posting is the 'preferred' *LaTeX document format*, introduced since arXiv's founding, which is technically demanding and is less likely to be mastered by non-scientists (Gunnarsdóttir 2005).[23] The regular introduction of additional controls over arXiv postings and the nature of the marshy ground indicate the problem of demarcation as it is encountered within physics and introduces the kind of problems we deal with here. Note that arXiv's criteria for exclusion are essentially social: they are ways of indicating whether the potential contributor is 'one of us'. arXiv´s criteria will form the subject of future studies by one or more of the current authors.

---

[20] Interview with Reyes-Galindo 16 June, 2014. These criteria can be compared with the discussion by Delborne (2008, p. 512).

[21] "The arXiv endorsement system", URL: http://arxiv.org/help/endorsement

[22] Interview with Reyes-Galindo 16 June, 2014.

[23] Though arXiv currently accepts other formats such as MS Word and HTML so the matter is somewhat more complicated.

**The marshy ground**

Part of the difficulty faced by arXiv is caused by the very characteristics of physics. For long periods the topic of 'foundations of quantum theory' was not thought of as belonging to respectable physics, with calculations and applications based upon measurements being taken to be the only appropriate way to move forward while the peculiar and counter-commonsensical interpretations were to be ignored. The work of such as David Bohm were discounted or wrongly dismissed by refereed journals within physics (Pinch 1977). These areas are also the focus of David Kaiser's 2011 book *How the Hippies Saved Physics* and his claim that the work of this quasi-fringe turned out to be of great value to such well-regarded modern subfields such as quantum computing and cryptography. The terrain of Foundations of Quantum Theory is also collaboratively shared by physicists and philosophers, something that at least some physicists would consider indicates something suspicious. The pendulum has swung back in recent times as an active physicist explained to us:

> Although I do not directly work on it myself, foundations of quantum physics has become very fashionable over the last 15-20 years or so. ... It is developing new experimental tests of the surprisingly non-intuitive implications of quantum physics [... and an] information theory approach gives new insight in trying to understand what quantum theory is, and is not, in the general class of physics theories in a rather abstract way. Quantum foundations is now a very large area internationally, and many mathematical physicists who were working in other areas in the 90s ... moved into this field; it is seen as a desirable field for many students choosing a PhD. In the UK, we have big Quantum Foundations groups in many places, including not only Imperial, but Oxford, Cambridge, Leeds, York, Glasgow and also Bristol. [24]

This area has, nevertheless, given rise to some of the most heated border disputes around postings on arXiv and the topic, according to some, reaches out to fields like parapsychology, with heterodox physicists developing theories based on the interaction between the physical world and consciousness which can be said to be justified by interpretations of quantum theory (starting with 'Schrödinger's Cat' whose state is said by some to be determined via the intervention of the consciousness of the observer – see Collins and Pinch (1982) for a simple introduction). [25] When Collins in 1972, and he with Trevor Pinch in the mid-1970s, spent time in California researching the sociology of the paranormal (Collins and Pinch 1979, 1982), the names and organisations that were frequently mentioned, linking physics with the 'consciousness movement', are still associated with arXiv's current border disputes. [26]

---

[24] Email from Mark Dennis to Collins 17 February, 2015.

[25] See also the discussion of the 'Quantum Coherence Heresy Group' in Collins 2004 p313ff.

[26] Foundations of quantum theory does not exhaust the marshy ground. Recently some string theorists have argued that their claims should not require observational support (Ellis and Silk, 2014), while solutions to physics problems such as the anthropic principle and the many-worlds hypothesis are speculative rather than observation based.

The names of these institutions can be found in the second line of Table 1, below: the Fundamental Fysiks Group, the Institute for Noetic Studies, and the Noetic Advanced Studies Institute. All three groups spring from the 1970s Californian intellectual milieu. The Fundamantal Fysiks Group can perhaps claim to have had the greatest impact on mainstream science. Formed at Berkeley in 1975 by Elizabeth Rauscher and George Weismann, the Fundamental Fysiks Group applied quantum physics to topics usually considered 'taboo' by mainstream science – telepathy, consciousness, remote viewing etc. The Fundamental Fysiks Group was largely made up of working physicists – other notable names include Fritjof Capra (author of The Tao of Physics, 1975), Wolf Prize winner John Clauser, and the sometime DARPA-funded unorthodox physicist Jack Sarfatti (who has supported arXiv dissenters, and has submitted to arXiv) – and several members of the group continued their work with government funding at the Stanford Research Institute (SRI).

More institutionalised than the Fundamental Fysiks Group is The Institute of Noetic Sciences (IONS), boasting several SRI alumni among its ranks, which was founded in 1973 by the astronaut Edgar Mitchell.[27] IONS' research on ESP, telekinesis, etc. often draws on quantum theory, hosting talks by scientists such as Russell Targ and Fred Alan Wolf. The Noetic Advanced Studies Institute (NASI) [28] has the greatest degree of interconnection with the other organisations described in this paper.[29] Founded in 1993 by Richard Amoroso, who claims a PhD in Cosmology and Philosophy of Mind from the (defunct) International Noetic University in California, NASI has 16 'senior fellows', including Rauscher and Francesco Fucilla, the founder of the Telesio-Galilei Academy of Sciences. Between 1998 and 2008 NASI published the *Noetic Journal*, which boasted an international editorial board with a handful of members affiliated to mainstream academic institutions, and between 1998 and 2011 it awarded the Noetic Medal of Consciousness and Brain Research. NASI is the current sponsor of the Vigier Symposia.

**The hinterlands of science**

The 'hinterlands' of physics proper begin beyond the marshy ground even though they have a presence within it and within arXiv's General Physics classification. Papers accepted in General Physics will be indistinguishable from other scientific papers in terms of their appearance and style but the same will apply to nearly all of those rejected entirely by arXiv. Many such rejected papers will be published in 'fringe' journals, or 'alternative' publication outlets.

The areas from General Physics outward are represented in Table 1 with Table 2 representing fringe outlets. Table 1 starts with arXiv's General Physics authors while, with some exceptions, the lower regions are more remote areas of the fringe (the left-right dimension having no significance). The second line of Table 1 shows organisations that are closest to the mainstream and occupy the marshy areas of foundations of quantum theory between the 'solid earth' of mainstream physics and the 'water' of the fringe with parapsychology mostly considered, by those on solid ground, to be entirely liquid. Parapsychology – that is 'scientific parapsychology' as practised in universities and the like – is included in this row because some consider that it has an overlap with foundations of quantum theory. The next row indicates organisations beyond the marshy area though many of those who inhabit this row will, at one time, have published in arXiv and mainstream journals and/or arXiv's General Physics category. The next rows are self-explanatory. There is some degree of internal sociometric connection within the top three rows but occupants of the remaining three rows are mostly disconnected. The shaded areas of Table 1 are fringe areas other than physics but we include them here to complete our classification of ways of being in the fringe and to indicate one direction in which future research on this topic will go. As already seen with parapsychology, the division between physics and non-physics is not sharp.

[27] "IONS Overview", URL: http://www.noetic.org/about/overview/

[28] "Noetic Advanced Studies Institute", URL: http://www.noeticadvancedstudies.us/

[29] However, in contrast to other fringe groups, NASI members do not appear to be interested in posting to arXiv.

| *arXiv General Physics population* | | | |
|---|---|---|---|
| Fundamental Fysiks Group | Noetic Advanced Studies Institute | Institute for Noetic Studies | Parapsychology |
| Natural Philosophy Alliance | Society for Interdisciplinary Studies | The Thunderbolts Project | Telesio-Galilei Academy |
| Natural Philosophy Society | Institute for Basic Research | Alpha Institute of Advanced Study | Common Sense Science |
| Creationism and intelligent design | Tobacco and oil lobby | Astrology | |
| Fringe archaeology and other fringe sciences | | | |

Table 1: Fringe institutions.

Table 2 shows most of the outlets where the occupants of Table 1 promulgate their ideas. For example, the paper discussed in Collins (2014) was published in one of the journals in line 2 of Table 2. General Physics, individual blogs and green ink letters – letters usually exhibiting unusual graphological or typographical conventions often sent to prominent scientist expressing a maverick theory of the universe or the like – appear on both tables as they are themselves outlets while also indicating locations in the terrain of the fringe.

| arXiv General Physics | | | |
|---|---|---|---|
| Vixra | Progress in Physics | Galilean Electrodynamics | Apeiron |
| Vigier Symposia Proceedings | General Science Journal | The Hadronic Journal | Parapsychology journals |
| Individual blogs | Green ink letters | Popular books | Newspapers |

Table 2: Fringe physics outlets

Figure 2 shows extracts from some of the kinds of materials mentioned in Tables 1 and 2.

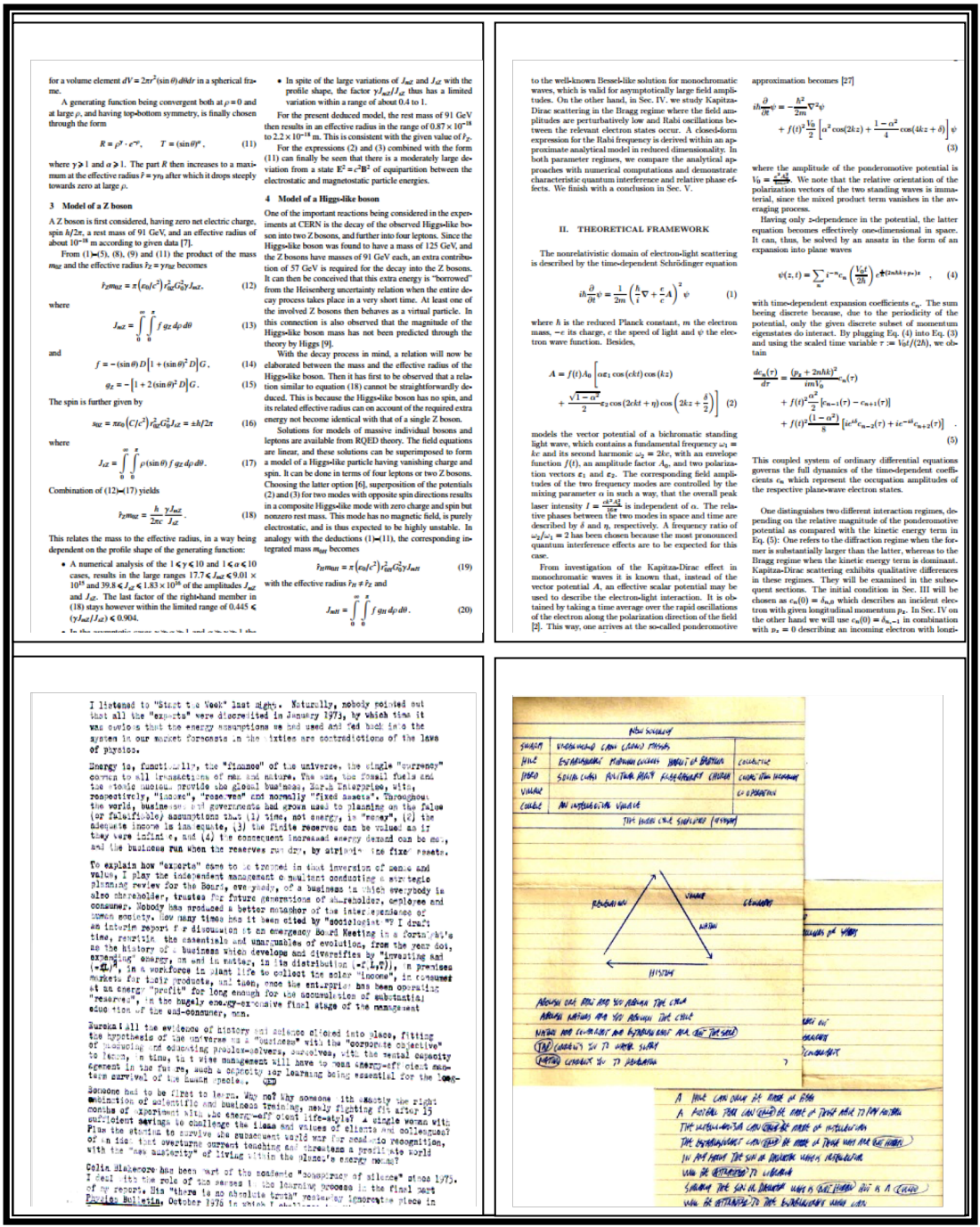

**Figure 1: Extracts from paper in arXiv, a paper in a fringe physics journal (not necessarily in that order) and two 'green ink letters'** [30]

---

[30] These materials are presented for their style not their content – they are meant to illustrate what such items look like – e.g. note similarity in mathematical content in the first two papers. Though the letters are addressed to an individual, we take them to be public documents intended to be promulgated (there will typically be many recipients). We understand them as intended in the spirit of letters to one's MP or Congressperson, or letters to a newspaper.

**The fringe as a community**

Some idea of the extent of the activities discussed here can be obtained from the 'World Science Database', which has now transformed into the 'Natural Philosophers Database' (NPD) associated with the Natural Philosophy Alliance and the breakaway Natural Philosophy Society (Table 1, row 3, left, and see below). The mission of the NPD 'is to catalogue all dissident science work world-wide in one place'.[31] In late summer 2014 there were 2290 people listed on the NPD, with 790 different interests listed on their profiles. The most common interests are: 'Relativity' (249), 'new energy' (211), 'gravity' (164), 'aether' (108 people), 'electric universe' (99), 'antigravity' (92), 'expanding Earth' (81) and 'cold fusion' (65).

Though we are going to treat the entries in Table 1 and Table 2 individually, a large proportion of the fringe has characteristics of a distinctive community. Members will often meet at the same conferences and organisers are interconnected. For example, Cynthia Kolb Whitney was the president of the Natural Philosophy Alliance (NPA) and editor of *Galilean Electrodynamics*, while William Lucas, who was the NPA vice-president, is also a scientist with Common Sense Science. Francesco Fucilla, the 'founding father' of the Telesio-Galilei Academy of Science (TGA) (which began as the Santilli-Galilei Academy, Ruggero Santilli being founder of the Institute of Basic Research), is also a fellow of the Noetic Advanced Studies Institute (NASI). NASI fellow Elizabeth Rauscher was a founder of the Fundamental Fysiks Group. Myron Evans, who heads the Alpha Institute for Advanced Studies (AIAS), helped establish the Vigier Symposia, which are now sponsored by NASI, while NASI founding director Richard Amoroso is also listed as member of the AIAS. The NPA's Sagnac Award has been given to Halton Arp (late editor of Apeiron) and Donald Scott, both luminaries among 'Electric Universe' theorists, while the TGA has awarded Gold Medals to, for example, Myron Evans and Wallace Thornhill, who is one of the founders of The Thunderbolts Project.

The sense of community, fragile though it may be, is also indicated by certain common characteristics not shared by mainstream science – we now describe three of these. We stress that we are *not* describing *these* characteristics as a way of marking off the fringe as a distinct socio-cognitive community – that exercise will come later. It is also vital to stress that these three characteristics only capture a proportion of the scientists and institutions described here. For example, a number are physicists still in mainstream jobs would be horrified to find themselves associated with those putting forward Jewish conspiracy theories.

Firstly, and unsurprisingly, the typical age of a contributor to the fringe seems to be considerably higher than that of contributors to mainstream science. According to the NPD,[32] the average age of those in the database is 65.1 and when Bartlett attended the 2014 NPA conference he was, at 37, by some margin the youngest presenter. By contrast, McWhinnie (2013) in a survey for the Institute of Physics, found that average age of permanent academic staff in the UK is 44.7 for men and 40.6 for women, not including contract researchers, who tend to be younger. Retired scientists with time to do unorthodox and unpaid work and whose career would no longer be put at risk by thinking outside the mainstream box seem to make up a disproportionate element of the community.

---

[31] "Natural Philosophers Database", URL: http://db.naturalphilosophy.org/

[32] "Natural Philosophers Database > Birthdays", URL: http://db.naturalphilosophy.org/birthdays/

Secondly, there is a surprising readiness to discuss the possibility that the resistance of the mainstream to fringe ideas is the consequence of mainstream cabals, particularly, a Jewish conspiracy. The website scientificethics.org, makes allegations of 'Jewish corruption' and 'scientific gangsterism'[33] as a cause of the 'persecut[ion] of the Italian American scientist R. M. Santilli', leading to the suppression of unorthodox scientific ideas, particularly those that conflict with 'organized Jewish interests on Einstein'.[34] A previous PhD research project at Cardiff on the danger of genetically manipulated organisms also ran into Jewish conspiracy theories when fringe ideas were rejected. Ethnographic research revealed that a casual, 'matter-of-fact', conversation can be held over dinner in certain fringe organizations about Einstein's one-time support for Israel leading the large number of Jews in modern physics to support relativity against all opposition. This is an uncomfortable echo of the 'Jewish physics' notions of the Nazis.[35] We wrote to a number of senior mainstream physicists asking if this kind of accusation ever reached their ears but they told us that this was no part of day-to-day physics nor had they ever heard of such things. Clearly this kind of idea violates the Mertonian norm of 'universality' and this might be said to distinguish at least a proportion of the group from the socio-cognitive activity of mainstream science. We will see that 'norm-violation' is one of our demarcation characteristics. But since much of the fringe does adhere to the Mertonian and other typically scientific norms, and since a proportion of the mainstream violates the norms, norm violation will be a useful criterion for policy-makers on only rare and specific occasions.[36]

Thirdly, there are a large number of engineers, particularly electrical engineers, populating the fringe. For example, of the 31 speakers at the EU2014 conference organized by the Thunderbolts Project, at least 11 had backgrounds in engineering, 5 of which were electrical engineers. We have seen above that this was noted by Michael Berry, former editor of the *Proceedings of the Royal Society*. We asked Berry why he thought this was and he remarked that he thought engineers' views were based in a 'practical working knowledge and a sound intuitive understanding of aspects of electromagnetism' but this, he believed, could make it hard to accept the counterintuitive consequences of a relativistic world view. Physicists and engineers are two different cultural groups, differently educated, but dealing successfully with the same phenomenon. The two traditions clash head-on only very rarely as in the case of the development of GPS. Here is how it looks from the viewpoint of a member of the one of the fringe groups:

---

[33] "Some of the scientific gangerisms perpoetrated[sic] by the Jewish physicist Steven Weinberg", URL: http://www.scientificethics.org/Steven-Weinberg.htm.

[34] "Documentation of Jewish Dishonesty and Corruption on Prof. Santilli's Article at Wikipedia", URL: http://www.scientificethics.org/Wikipedia-corruption.htm. This website comes very close to being anti-Semitic, for example explaining Hitler's actions by reference to the controlling influence of Jews in Germany.

[35] In 1938, *Nature* published an anti-Jewish-physics letter from an institutionally powerful German Professor of physics (Stark 1938). See also Wazeck (2014).

[36] The norms are explored in greater depth in a work on the relationship between science and society (Collins and Evans, *Why Democracies Need Science*, forthcoming).

[T]here are quite a few with engineering backgrounds in the NPA. ... I think this has to do with engineers being more pragmatic, and not attached to any particular ideology. Particularly in theoretical physics, much of the "standard" science is based on a set of shared assumptions ... or perhaps the better word is "dogma". Engineers are typically practical people, who use science and apply it to the real world. Since their careers are not anchored to these dogma, they don't have any problems questioning the Big Bang [for example]. Another example is when the first GPS satellites went up, all the engineers thought Einstein's concept of warped space-time was nonsense. It turns out that some of these engineers became believers since the Einstein-based adjustments proved to work. However, some engineers and scientists in the NPA have demonstrated that Lorentz transformations work equally well for GPS satellite adjustments ... I also have a background in electrical engineering. I look for new physics which I can apply. And when I come across any physics, my ultimate arbitrator is "can I build something with this, or can I verify it through experiment". I won't be able to build an anti-gravity machine based on current physics. I also won't be able to build a device for superluminal (faster than the speed of light) communications which according to Einstein, would be impossible. But I don't really care what Einstein said … My career does not depend on whether I am an Einstein "believer" or not.[37]

**Analytic description of the hinterlands of science**

We now show how the institutions and outlets listed in Tables 1 and 2 differ as socio-cognitive enterprises from mainstream science in terms of, among other things, 'formative intentions' (Collins and Kusch, 1998). Formative intentions are what drive the actions that that members of cultural groups aspire to and which give rise to their characteristic 'form of life' (Winch 1958; Wittgenstein 1953) – taking out mortgages in some societies, divining witches in others, and so forth. Our demarcation criteria turn, in part, on distinguishing the formative intentions characteristic of the fringe from those that characterise the mainstream.

We organise the discussion of different institutions under eight characteristics. Two of these characteristics represent the default position of mainstream science: the coherence and authority on the one hand and the individualism on the other that constitute 'the essential tension'. The other six characteristics are discussed under separate sub-heads. Three of the six differences we will draw out come from previous work and three are new.[38] The discussion works roughly from-out-to-in, in terms of the geography of the fringe. The schema is set out in Table 3, where the characteristics are found in the top row numbered 1-8 with exemplifying activities in the left hand column. A shaded square indicates that the corresponding activity is characterised by the label in the top row. The presence of shaded squares in columns 3 onwards indicates differences with mainstream science. A square with a heavy border but no shading indicates that the activity may or may not exhibit the characteristic. The top-left to bottom-right direction corresponds roughly to distinctiveness – or distance under our ecology metaphor – from mainstream science. We work through the characteristics and will then go back and classify the institutions that have not already been used as examples.

---

[37] E-mail communication to Bartlett 30th January 2015.

[38] By nature of this kind of analysis, which deals with family resemblance rather than necessary and sufficient conditions, boundaries are fuzzy and the way we assign activities into them is open to debate.

*Oblique Orientation*

The first difference between certain fringe activities and mainstream science is *oblique orientation* (column 8). Indeed, it is not clear if activities thus characterised should be counted as candidates for the label science in the first place. Astrology, the typical activity presented, is not really aiming to be a science – at least not a contemporary science – but to appeal to the public rather than other scientists. Fringe archaeology – supposed remnants associated with Arthurian legends and the like – also fits here because it is directed more at providing material for popular books than at impacting on the science of archaeology. Perhaps something similar could be said for the works of Erich von Daniken and other books positing past cosmic catastrophes or visits from aliens and the like. *Conspiracy theories* – 'the Moon landings were faked; corpses of aliens can be found at Roswell' – are also not intended to be absorbed into mainstream science.

| Characteristics of the activity / Exemplifying sciences | 1 Coherent and Authoritarian | 2 Individualism | 3 Past sell-by date | 4 Pathological Individualism | 5 Primarily oppositional | 6 Revolutionary intent | 7 Counterfeit or norm-violating | 8 Oblique Orientation |
|---|---|---|---|---|---|---|---|---|
| Mainstream science | ██ | ██ | | | | | | |
| Foundations of quantum theory | ██ | ██ | | | | | | |
| Parapsychology | ▒ | ▒ | ▒ | | | | | |
| Natural Philosophy Alliance (NPA) | | | | ▒ | ▒ | ▒ | | |
| Individual blogs Green Ink Letters | | | | ▒ | ▒ | | | |
| Creationism and Intelligent Design | ▒ | | | | ▒ | ▒ | | |
| Tobacco and oil lobby etc | | | | | ▒ | | ▒ | |
| Astrology | | | | | | | | ▒ |

Figure 2: Ways of being in science.

*Norm-violation*

Column 7 represents *norm-violation* by which we mean the so-called 'research' purchased by the tobacco and oil lobbies so as to fabricate doubt and create 'counterfeit scientific controversies'.[39] Jewish-conspiracy theories would also fit here but not in a policy-useful way.

*Revolutionary Intent*

The next outermost characteristic (column 6) is *revolutionary intent*. Collins and Evans (2007) claim that for something to be counted as science – say, Joe Weber's defeated gravitational wave claims (Collins 2004) – the author of the claim should be aiming to preserve as much of existing science as possible. If the work is revolutionary they should be reluctant revolutionaries aiming to change as few concepts, empirical assumptions and experimental procedures as possible.[40] If the aim is radical transformation of the institution of science then what is being done is not continuous with science. Thus, though Einstein engendered a revolution in our understanding of space and time, and Joseph Weber could have given rise to a revolution if his findings had been believed and interpreted in certain ways (a revolution and a proto-revolution in the Kuhnian sense), this was not their authors' primary intent. Einstein and Weber wanted to preserve the existing observational nature of science.[41] Creationism and intelligent design, in contrast, would involve a shift in the order of what counts as evidence, raising the contents of old books of obscure origin to a much higher status within science when it comes to observation-based claims.[42] The formative intentions of creationism therefore differ from those of mainstream science in a marked way and this puts us in a position to say to policy-makers: 'in respect of understanding scientific consensus, you can ignore any claims arising out of *creationism* because they are not claims arising out of an institution that is continuous with science even though their proponents direct them toward science.'

*Sell-by-date*

Columns 4 and 5 are central to the fringe and we will look first at column 3. For a claim to be 'past its sell-by date' means that science as a social organisation has allowed any controversy associated with the claim to drift out of focus, even though it was once a topic of hot debate in the mainstream.[43] The proponents of the idea are likely to be able to point out that it has not been thoroughly defeated by logic or observation and refuse the characterisation. Parapsychology, that is to say, 'scientific parapsychology', as conducted in university departments and the like, with its careful, statistically-analysed experiments and peer-reviewed journals, is indistinguishable from mainstream science in terms of the eight characteristics except that it has been making claims for so long without any breakthrough success that it has ceased to be a matter of concern to the mainstream. It is now mainly criticised by philosophers, stage-magicians and the amateurish 'skeptics' movement rather than by scientists: it is past its sell-by date.[44] The difficulty with this category is that all anti-relativity movements, for example, could also be said to be past their sell-by date as the principal opposition to the theory of relativity faded away some decades past. We choose to treat these movements as primarily oppositional, however, reserving the sell-by-date criterion for movements which are not essentially oppositional: parapsychology is not against anything but, rather, wants to add an extra dimension to existing science. Elsewhere we applied the concept to unorthodox theories of HIV. These have a more oppositional character but it could be said that this is not the principal motivation but a consequence of the alternative view.[45]

---

[39] Oreskes and Conway (2010); Collins and Weinel (2011).

[40] The same demarcation criterion was put forth by Lakatos (1978).

[41] After writing this passage we discovered that Max Planck was also a reluctant revolutionary (Kragh 2000).

[42] Though intelligent design appears to work independently of old and obscure books, its unfalsifiable hypothesis which leads to no new avenues of research would not be posited without the influence of such sources.

[43] Collins and Weinel (2011).

*Primarily Oppositional and Pathological Individualism*

Returning to column 5, *primarily oppositional*, indicates opposition to mainstream science as the main organising principle of the activity. The prime goal of those pursuing such an enterprise is not to advance science but to oppose certain of the findings of existing science – for example, the category includes those whose main aim is to find flaws in the theory of relativity. Of course, 'organised scepticism' is a feature of regular science but this refers to specific results rather than the mainstream. This characteristic also tends to be closely related to *pathological individualism* (column 4), which refers back to the essential tension. We take it to be a characteristic of science as we know it that there is always a tension between authority and coherence on the one hand and individual brilliance and heterodox discovery on the other. In fact, in terms of the categories listed in Table 3, the essential tension, along with the absence of characteristics 3-8, is what defines mainstream science. As we can see, under this scheme, the 'marshy ground' of quantum foundations counts as mainstream science even though many of its practitioners have found themselves pushed into arXiv's General Physics category or have been excluded from arXiv altogether.[46] Pathological individualism is exhibited when the *main concern* is with individual, heterodox, brilliance without any recognition that it has to be in tension with coherence and authority. This is *pathological* individualism, absence from column 1 implying presence in column 4. Green Ink letters and individual blogs exhibit pathological individualism because there is no peer review or community assessment before promulgation.[47]

---

[44] The notion of sell-by-date could, with a stretch, be said to have a resonance with Langmuir's (1989), 'The ratio of supporters to critics rises up to somewhere near 50% and then falls gradually to oblivion,' though Langmuir does not see the phenomenon as sociological.

[45] Weinel (2007, 2010).

[46] Which is simply to say that here demarcation criteria are more subtle.

The Natural Philosophy Alliance (NPA) is one of the most active, diverse, and well populated organisations on the fringe and is a paradigm of pathological individualism and oppositional stance. A statement by one of its leaders sums up its approach:

> Instead of trying to play the consensus game … we're going to be like everything else: in [the] arts, could you imagine if everyone paints the same? … Consensus is not only wrong, but detrimental and dangerous. It keeps us from true scientific progress. (David de Hilster at the 19[th] Annual Natural Philosophy Alliance Conference, 2012, Albuquerque, NM.[48])

Since this group exhibits this characteristic so clearly, we will spend extra time describing it and closely related institutions. Formed in 1994, by summer 2014 the NPA claimed just over 130 paid-up members and more than 800 'members' on its website, though in summer 2014 the NPA split, with a two organisations emerging, the Natural Philosophy Alliance and the John Chappell Natural Philosophy Society (NPS)[49]. While the organising committees and other administrative organs of these two associations are in conflict, there is significant overlap in membership. Neither the NPA or the NPS have its own journal, but publishes the proceedings of its annual conference electronically and in print using self-publication services such as Lulu. it also provides links to electronic versions of the papers of 'members' (not always with their consent or foreknowledge).[50] The NPA has organised 21 Annual Conferences (the most recent organised by the post-split NPA, with the NPS holding its inaugural conference in August 2015). Running over several days, they draw in a wide range of fringe physicists and include the presentation of the 'Sagnac Award'. The *Proceedings* for the 2013 Annual Conference, the last before the split, run to nearly 400 pages. Even the diminished, post-split 2014 NPA Annual Conference, at which Bartlett presented a paper and conducted ethnographic observation, attracted delegates from the UK, Australia, Colombia, New Zealand, and Russia. Of the 23 presenters listed on the programme, at least nine have doctoral-level degrees in physics or a related subject, with a handful holding academic positions in universities.

---

[47] An example of extreme individualism was mathematician Grigori Perelman's unexpected posting on arXiv of the critical missing steps to prove the Poincaré conjecture, which had eluded the world's greatest mathematicians. The proof was not submitted for peer review by Perelman, although it was later verified by mathematicians. Perelman retired from mathematics and declined the prizes and honours associated with the proof, including the 1 million dollar Clay Millennium Prize and the Fields Medal. The sociological notion of 'pathological individualism', however, applies not to individuals but to institutions so what Perelman did, though very unusual, could not be counted as pathological individualism. Though green-ink letters and individual blogs are promulgated by individuals it is the institution which they constitute which we count as pathologically individualistic.

[48] Video available at "Consensus in Science is Wrong", URL: https://youtu.be/UABe5oiYUCU

[49] "John Chappell Natural Philosophy Society > About", URL: http://www.naturalphilosophy.org/site/about/

[50] As reported to Bartlett during fieldwork.

The Natural Philosophy Alliance and the John Chappell National Philosophy Society adopt some characteristics of mainstream scientific institutions, but they also differ in crucial respects. On the one hand the these organisations sometimes appear to be 'science orientated', as illustrated when a speaker at the 2014 NPA Annual Conference presented material that implied support for 'Young Earth' Creationism[51]. In a heated e-mail exchange immediately after the conference, several members of the NPA voiced objections on the basis that this was non-scientific in that it was derived from a reading of the Bible rather than observation and experiment. On the other hand, these organisations espouse a strongly individualistic model of science that makes such boundary work difficult: 'The NPA wants to change the current philosophy of science and return to the ancient Greek approach to natural philosophy based on the logical approach of the axiomatic method.'

The NPA and the NPS are therefore primarily organised around its opposition to mainstream physics. The pre-split NPA webpage listed the problems of contemporary science: 'The Big Bang theory is fundamentally flawed […] Relativity has flawed assumption and when proof for such is examined, it is not proof at all […] Expansion tectonics (the earth is expanding / growing) is a much better model than modern-day plate tectonics […] The universe is way more electrical than currently thought […] Most all NPA scientists agree that science took a huge wrong turn in the early part of the 20th century'.[52] In the literature of the NPA we also find a perfect expression of pathological individualism:

> Following the words of Galileo "In questions of science, the authority of a thousand is not worth the humble reasoning of a single individual", the NPA does not accept any authorities in science except logic and empirical data. Science is not a democratic process. Just as the world would have benefited from listening to the words of Galileo during his lifetime, the NPA champions the right and necessity of all natural philosophers to be given a fair hearing based on the logical and experimental basis of their work instead of its "political correctness" under the current philosophy of science.[53]

---

[51] While there are several flavours of creationism, Young Earth Creationism is the religious belief that the Earth was created mere thousands of years ago (typically about 6000) by God. While this is by no means a popular idea on the fringes of physics examined in this paper, several theories espoused by fringe physicists – for example the idea that the speed of light has been slowing (and other 'constants' have also be changing) have been slowing - have been deployed to support an Earth much younger than is held by mainstream science. Some of this work has touched on the institutional 'marshy ground' – see, for example Norman and Setterfield (1987) *The Atomic Constants, Light and Time*, an invited report published by Stanford Research Institute (which was, among other things, an institutional refuge for members of the Fundamental Fysiks Group) and Flinders University in Australia.

[52] "Problems in Mainstream Science", URL:
https://web.archive.org/web/20130729040741/http://www.worldnpa.org/site/problems-in-mainstream-science/

[53] "Principles of the Natural Philosophy Alliance" URL: http://worldnpa.org/about/principles/

Inasmuch as these organisations hold a model of science, it is that progress comes through the iconoclasm of individuals overturning stale orthodoxies:

> We value free expression and vigorous debate of scientific thought; and we reject the assertion that scientific validity may be established through consensus[54]

These are ideal expressions of one side of the essential tension; that the authority of coherence of belief and consensus is, by its very nature, suspect:

> **Science in the mainstream is dominated by politics, not science.** Criticism of Einstein, the big bang, and other mainstream theories is not allowed in the mainstream whereas in all other human endeavours including music, art, literature, business, politics and engineering, opposing ideas are necessary for coming up with the best solutions humans can muster. The NPA encourages diverse opinion, believing that better truths will emerge.
> **Most all NPA scientists agree that science took a huge wrong-turn in the early part of the 20th century.** Many NPA members independently and collectively are starting physics and cosmology over from the time of Einstein in 1905 in order to put science back on track.[55] [original emphases]

The result of this kind of approach to science is reflected in the cognitive and social life of the fringe as a whole in that organisations are continually splitting and reforming with bitter disputes turning on the sets of ideas of individuals; many of the organisations are associated with named individuals in a way that the organisations of mainstream science are not. The phenomenon was observed within the organisations by Bartlett during the 2014 NPA conference in Baltimore and the 2015 NPS conference in Boca Raton. Both were, on the face of it a scientific conference, but they were loosely organised administratively – at times disorganised – and without cognitive coherence. The delegates were brought together by their opposition to the mainstream, with each delegate expressing their opposition in their own terms – an expanding earth, an electric universe, an eternal and evolving universe. The NPA and NPS conferences were a space for the presentation of any number of different ways of being in opposition to the mainstream. The tendency to schism among the organisations and the administrative disorganisation is a nice example of homology between cognitive and administrative organisation. Each of the scientists, cherishing individuality, distrusts authority and organisation. Any residual unity is not brought about by sharing a common goal other than to be against the mainstream. Members of the organisation themselves recognise the problem:

---

[54] " Mission Statement", URL:
https://web.archive.org/web/20130801162841/http://www.worldnpa.org/site/mission-statement/
[55] "Problems in Mainstream Science", URL:
https://web.archive.org/web/20130729040741/http://www.worldnpa.org/site/problems-in-mainstream-science/

The alternative to the mainstream has not been organised. So what you find is most of them start off believing the mainstream, they fall out of the mainstream, and they think they've got something new to revolutionise physics. So they all come away believing they're the new Einstein. So they're the ones forming part of the NPA. Everybody talks to each other, they've got their own pet theories, but they're not going to get organised around one theory because they're all promoting their own pet theory. And that's what the mainstream has created. It's disorganised the alternative. The alternative to the mainstream has not been able to organise, because the voice has been silenced.[56]

This degree of individualism has to be 'pathological' because the NPA's claim – 'the NPA does not accept any authorities in science except logic and empirical data'— would imply that no-one else's observations or calculations could be ever trusted. In this scenario, science as we know it would grind to a halt – trust goes with social organisation and social coherence. We can, then, advise policy-makers: if a group is driven primarily by opposition to the mainstream then there are grounds for taking their views less seriously.

**Remaining institutions**

We now provide brief sketches of the remaining institutions, indicating which of the characteristic they share and where they would fit on Table 3. These remaining institutions will be found in rows 2 and 3 of Table 1 and row 2 of Table 2.

*The Thunderbolts Project* was founded in 2004 to promote the Electric Universe (EU) paradigm, which 'emphasizes the role of electricity in space and shows the negligible contribution of gravity in cosmic events.'[57] The Thunderbolts Project holds annual conferences and publishes books and DVDs through Mikimar Press. As with the *Society for Interdisciplinary Studies* (below), there is a strong Velikovskian strain to their membership, and the contribution of electrical engineers is evident. Bartlett attended are participated in the EU2015 – Paths of Discovery conference in Phoenix. These annual conferences have grown to quite large events, with over 200 attendees and a similar number subscribing to a live stream of the presentations. Unlike the NPA and the NPS, *The Thunderbolts Project* avoids pathological individualism because of the unity of its ideas. One illustration of this is the attendance at EU conferences of an *audience* comprising many non-presenting attendees. Another is the development of a collaborative experimental programme into the 'electric sun' hypothesis, supported with funding from the International Science Foundation[58]. The intention to explain the whole universe in terms of electrical forces would not count as 'revolutionary' in our terms though. Were it to succeed, it would certainly be a Kuhnian revolution, albeit a backward looking one. The emphasis on historical and mythic records of catastrophe, popular books, and the insights of non-scientists may, however, place the organisation in the following columns: 8, because it has an element of oblique orientation; 6, because if successful it would precipitate an institutional revolution in science; and, perhaps, 5, because opposition is a strong driving force.

---

[56] Interview with Bartlett, 26th June 2014.

[57] "Exploring the Electric Universe", URL: https://www.thunderbolts.info/wp/about/syn/

[58] SAFIRE (Stellar Atmospheric Function in Regulation Experiment) appears to be the only project currently supported by the International Science Foundation (ISF), Funding for SAFIRE is described in this way: "Through private funding, ISF offered $1,000,000 for its initial financing with $1,200,000 for continued funding through 2015". "SAFIRE Project", URL: http://isciencefoundation.org/safire/

*The Society for Interdisciplinary Studies* was founded in 1974, another organisation inspired by Immanuel Velikovsky. The SIS straddles fringe archaeology and fringe physics. Presenters at their meetings discuss catastrophist interpretations of pre-history alongside ideas such as the Electric Universe. Evidence of ancient catastrophes play a central role in their work and they claim that '[many] great discoveries and insights are made by intuitive non-scientists'.[59] The analysis of this group in terms of characteristics is similar to that of *The Thunderbolts Project* (above).

Founded in 2008, the *Telesio-Galilei Academy of Science* is not overtly oppositional, but 'champions the true scientific spirit, promoting 'courageous departures from conventional perceptions', citing Copernicus, Newton and Einstein.[60] The Annual Gold Medals that it awards in this spirit are, most often, given to fringe physicists. It is certainly characterised by pathological individualism as Galileo's name in its title indicates. Galileo is a frequently recurring motif in the fringe.

*The Institute for Basic Research* was founded in 1981 by Ruggero Santilli to promote his 'Hadronic Mechanics'. The IBR listed its membership at 135 in 2008, claimed 'scientific addresses' six countries, and controls the Hadronic Press, which publishes two journals and a number of monographs. Once more, the substance of the IBR's program is more directed at a Kuhnian rather than an institutional revolution but the readiness with which it's supporters endorse the idea of a Jewish conspiracy could class it as having revolutionary intent and being norm violating. Its strong leadership style suggests pathological individualism and an emphasis on opposition to mainstream science.

*Alpha Institute for Advanced Study* was founded in 1998 to promote Myron Evans' Einstein-Cartan-Evans (ECE) theory – a Unified Field Theory, which refutes, 'in great detail' '[n]early all the precepts of standard physics'.[61] The AIAS claims 38 named 'fellows' in addition to Evans. The AIAS expressly rejects standard forms of scientific dissemination, preferring self-publication on the AIAS website claiming that journal publication is obsolete and restrictive while the web allows measures of actual usage to indicate significance.[62] It is therefore pathologically individualistic, and primarily oppositional as well as having revolutionary intent in terms its views on publication.

---

[59] "SIS Background", URL: http://www.sis-group.org.uk/sis-background.htm

[60] "Telesio - Galilei Academy's Mission", URL: http://www.telesio-galilei.com/tg/index.php/missions-and-goals

[61] "Alpha Institute for Advanced Studies (AIAS)", URL: http://www.aias.us/

[62] "Overview of ECE Theory", URL: http://www.aias.us/index.php?goto=showPageByTitle&pageTitle=Overview_of_ECE_Theory

*Common Sense Science* is a small group, seemingly based around three or four physicists and or electrical engineers who publish a newsletter and journal, *Foundations of Science*, along with privately distributed books and videos. They pursue a physics that reflects the 'Judeo-Christian Worldview' and are opposed to 'quantum reality, randomness, and multiplicity of force laws'.[63] CSS seems unified but they also seem primarily oppositional and, given the central role that religion plays in their cognitive lives and institutional activities, they appear to have revolutionary intent.

**Remaining outlets**

We now move to row 2 of Table 2, the remaining outlets that correspond to the institutions described above,

*viXra* (arXiv spelled backwards) is an electronic preprint server founded in 2005, by Phil Gibbs, a former physicist. viXra has over 10,000 papers in its archive, and claims to be 'truly open', accepting all submissions except those that are 'vulgar, libellous, plagiaristic or dangerously misleading'.[64] It appears to have a reaction to arXiv's rejection of various classes of papers. Given its degree of openness it would seem to have revolutionary implications in respect of publication practices in science and to encourage pathological individualism.

Founded in 2005, *Progress in Physics* promotes 'individual academic freedom and will consider all work without regard to affiliations'.[65] It published a Declaration of Academic Freedom[66], arguing that peer review had become a tool of censorship so it has revolutionary intent and encourages pathological individualism though the papers it publishes have all the appearance of scientific papers so it does not operate without any constraints.

*Apeiron*, published between 1987 and 2012, was closely associated with Halton Arp, a critic of Big Bang Theory. The editorial board was made up of scientists and mathematicians holding academic positions, but published work from fringe physicists associated with the NPA, and the TGA. We do not know enough about *Apeiron* to classify it with confidence.

*Galilean Electrodynamics* (GED) is a journal, founded in 1989 by Petr Beckmann (d. 1993), a professor of Electrical Engineering, to promote his anti-relativity position. The current editor is Cynthia Kolb Whitney, vice-president of the NPA. The journal is avowedly oppositional, describing itself as 'devoted to publishing high quality scientific papers, refereed by professional scientists, that are critical of Special Relativity, General Relativity, Quantum Mechanics, Big Bang theory and other establishment doctrines.'[67]

---

[63] "Worldview Principles", URL:
http://www.commonsensescience.org/worldview_principles.html

[64] "Why viXra?", URL: http://vixra.org/why

[65] "Progress in Physics", URL: http://www.ptep-online.com/

[66] "Declaration of Academic Freedom", URL: http://www.ptep-online.com/index_files/rights.html

[67] "Galilean Electrodynamics", URL:
http://web.archive.org/web/20151002140914/http://home.comcast.net/~adring/

The *Vigier Symposia* have been running since 2005. The Vigier Symposia are sponsored by the Noetic Advanced Studies Institute, an institution located in the 'marshy ground'. The symposia are well embedded in the fringe of physics, with, for example, Telesio Galileo Academic gold medals awarded alongside the symposia in 2010 and the 2014 symposia being well attended by Natural Philosophy Alliance members. Scientists with current academic positions also present at these conferences, however, and the proceedings have been published by such mainstream publishers as the American Institute of Physics, Springer, and World Scientific. Therefore the Vigier Symposia seems to straddle the marshy ground.

The *General Science Journal*, while starting as a home for criticisms of Special Relativity, provides an outlet for pathological individualism as well as implying revolutionary intent in respect of publication practices. The journal sees itself as, 'provid[ing] an opportunity for public presentation of scientific theories without prior and arbitrary assessment, criticism or rejection by the recipient. Judgement by the few runs counter to the spirit of scientific exploration' [68]

**Conclusion**

In this paper we have asked and tried to answer a question that has arisen out of the revolution in the social studies of science that took place in the 1970s (and has been referred to as 'the second wave of science studies'). We look at problems for social scientists and decision-makers that arise out of the proscription on simply drawing on scientists' authority when it comes to judgments of scientific truth. We look for sociological ways to understand the difference between mainstream the fringe science; we try to describe differences in the form of life of the mainstream and the fringe. Scientists are already more than capable of handling science for themselves but science studies can help enrich and legitimate the decision-making process. We have found a number of analysts' demarcation criteria based in the form of life of science, a paradigmatic example being a difference in the balance of 'the essential tension' in the case of the fringe compared to the mainstream. What we believe we have done is to make just a start on understanding how science studies could provide a better understanding of the relationship between science and technological decision-making that is based in a social understanding of the institution of science.[69] The more difficult task will be to extend this kind of analysis to what we have called 'the marshy ground.'

We also address a long-standing question for the authors: what is the model of science that informs the writers of green-ink letters and other members of the fringe? The answer is one in which isolated individuals are often best placed to plumb the secrets of nature while consensus is dangerously authoritative. It is an idealised, a-social, model of science. Strangely, it appears as if science's idealised model of itself as a kind of logical machine is precisely what gives rise to the fringe; many members of the fringe believe they are the only true upholders of this asocial model. Strangely, to the extent to which science studies scholars see their role as supporting the claims of the fringe and including them in a levelled out science, they are promulgating the very individualistic model of science which the 1970s revolution was supposed to have done away with.

---

[68] "Dedicated to the Free Expression of Scientific Theories", URL: http://gsjournal.net/Science-Journal/purpose

[69] See also Collins and Evans (2017, forthcoming)